\documentclass[12pt]{article}
\usepackage{epsfig,cite}
\usepackage{amsmath}
\usepackage{amssymb}
\usepackage{amsfonts}
\usepackage{latexsym}
\usepackage{wasysym}
\usepackage{bm}
\def\bea{\begin{eqnarray}}
\def\eea{\end{eqnarray}}
\def\beq{\begin{equation}}
\def\eeq{\end{equation}}
\catcode`@=11 
\def\slash#1{\mathord{\mathpalette\c@ncel#1}}
 \def\c@ncel#1#2{\ooalign{$\hfil#1\mkern1mu/\hfil$\crcr$#1#2$}}
\def\lsim{\mathrel{\mathpalette\@versim<}}
\def\gsim{\mathrel{\mathpalette\@versim>}}
 \def\@versim#1#2{\lower0.2ex\vbox{\baselineskip\z@skip\lineskip\z@skip
       \lineskiplimit\z@\ialign{$\m@th#1\hfil##$\crcr#2\crcr\sim\crcr}}}
\catcode`@=12 
\def\twiddles#1{\mathrel{\mathop{\sim}\limits_
                        {\scriptscriptstyle {#1}}}}
\newcommand\sss{\scriptscriptstyle}

\newcommand{\abs}[1]{\left|\,#1\,\right|}
\def\({\left(}
\def\){\right)}
\def\[{\left[}
\def\]{\right]}

\def\bsigma{\overline{\Sigma}}

\def\lb{\bar{L}}

\def\aq{\alpha_s(Q^2)}
\def\amu{\alpha_s(\mu^2)}
\def\ab{\bar\alpha}
\def\a{\alpha_s}
\relax

\def\abs#1{\left|#1\right|}

\def    \hepph  #1 {{\tt hep-ph/#1}}
\def    \hepex  #1 {{\tt hep-ex/#1}}
\def\qt{q_{\sss T}}
\def\qtq{q_{\sss T}^2}
\def\hqt{\hat q_{\sss T}}
\def\hqtq{\hat q_{\sss T}^2}
\def\qtvec{\vec q_{\sss T}}
\def\bvec{\vec b}
\def\abs#1{\left|#1\right|}

\newcommand{\sect}[1]{\setcounter{equation}{0}\section{#1}}

\newsavebox\tmpfig

\begin{document}

\pagestyle{empty}

\begin{flushright}

IFUM-905-FT\\CERN-PH-TH/2008-149\\ GeF/TH/3-08\\
\end{flushright}

\begin{center}
\vspace*{0.5cm}
{\Large \bf Borel resummation\\ of 
transverse momentum distributions}
 \\
\vspace*{1.5cm}
Marco Bonvini,$^{a}$ Stefano~Forte$^{b}$ and Giovanni~Ridolfi$^{a,c}$
\\
\vspace{0.6cm}  {\it
{}$^a$Dipartimento di Fisica, Universit\`a di Genova and
INFN, Sezione di Genova,\\
Via Dodecaneso 33, I-16146 Genova, Italy\\ 
\medskip
{}$^b$Dipartimento di Fisica, Universit\`a di Milano and
INFN, Sezione di Milano,\\
Via Celoria 16, I-20133 Milano, Italy\\
\medskip
{}$^c$CERN, PH Department, TH Unit,
CH 1211 Geneva 23, Switzerland}\\
\vspace*{1.5cm}

{\bf Abstract}
\end{center}

\noindent
We present a new prescription for the resummation of 
contributions due to soft gluon emission to the transverse
momentum distribution of processes such as Drell-Yan production
in hadronic collisions.  We show that familiar difficulties in obtaining
resummed results as a 
function of transverse momentum  starting from impact-parameter
space
resummation are related to the divergence of the perturbative
expansion of the momentum-space result. We construct a resummed
expression by Borel resummation of this divergent series, removing
the divergence in  the Borel inversion through the inclusion of a
suitable higher twist term. The ensuing resummation prescription is free of
numerical instabilities, is stable upon the inclusion of subleading
terms, and the original  divergent
perturbative series is asymptotic to it. We compare our results to
those obtained using alternative prescriptions, and discuss the
ambiguities related to the resummation procedure.

\vspace*{1cm}

\vfill
\noindent

\begin{flushleft} July 2008 \end{flushleft}
\eject

\setcounter{page}{1} \pagestyle{plain}

\sect{Transverse momentum resummation}

The computation of transverse momentum distributions of heavy systems
(such as  dileptons, vectors bosons, Higgs) plays an important role in
collider phenomenology, from the Tevatron to the
LHC~\cite{tev4lhc,higgs}. As is well known,
the perturbative QCD expansion of the inclusive distribution contains
to all orders powers of $\a \ln^2(\qt/Q)$, due to the
emission of soft and collinear gluons. When
 the transverse
momentum $\qt$ is much smaller than the mass of the final state $Q$
these logs become large and must be resummed in order for perturbative
predictions to remain reliable.

The resummation, to given logarithmic accuracy, can be
performed~\cite{CSS} for the Fourier transform of the differential
cross-section $\frac{d\sigma}{d\qt^2}$ with respect to $\qt$. Upon
Fourier transformation, $\qt$ turns into its Fourier conjugate, the
impact parameter $b$, and large logs of $\qt/Q$ become large logs of
$bQ$. Fourier transformation is necessary in order for the
contributions included by resummation to respect transverse momentum
conservation, thereby avoiding the spurious factorial growth of
resummed coefficients~\cite{cmnt}.  However, the Fourier transform
must be inverted in order to obtain resummed predictions for physical
observables. This is problematic because the Fourier inversion
integral necessarily involves an integration over the region of impact
parameters where the strong coupling is not well defined because of
the Landau pole.

This problem has been treated with various prescriptions. One
possibility is to modify the behaviour of the strong coupling in the
infrared in the Fourier inversion integral~\cite{CSS} ($b_\star$
prescription, henceforth): this procedure is widely used, but it is
known to lead to numerical instabilities when the resummed results are
matched to fixed--order ones~\cite{EV}.  A second option is based on
the observation that the Fourier inversion integral can be computed
order by order in an expansion of the resummed results in powers of
$\a$: if only leading log terms are retained in the Fourier inversion,
the result is then well defined for all values of
$\qt$~\cite{EV}. This procedure however is unstable to the inclusion
of subleading corrections: the Fourier inversion can be performed to
next-to-leading log accuracy~\cite{FNR} (as it is necessary if the
resummation is performed to this order), but in such case the result
differs significantly from the leading log one, and in fact for $Q$
around 100~GeV it blows up for values of $\qt$ of order of several
GeV, well within the perturbative region. A ``minimal'' prescription
which is free of these difficulties can be constructed~\cite{MP},
along the lines of the similar prescription for threshold
resummation~\cite{cmnt}. Namely, the integration path in the Fourier
inversion is deformed in such a way as to leave unchanged the result
to any finite perturbative order, but avoiding the Landau pole and
associate cut in the resummed result. This leads to a prescription
which is free of numerical and perturbative instabilities: its only
shortcoming is that it is difficult to assess the ambiguities related
to the resummation procedure, as it can be done in the $b_\star$
prescription by varying the way in which the infrared behaviour of the
strong coupling is modified.

Here we shall show that, analogously to what happens in the case of
threshold resummation~\cite{frru}, the ambiguity in the resummation
procedure is due to the fact that the perturbative expansion of the
resummed result for the transverse momentum distribution itself in
powers of $\a$ diverges.  After discussing, in the next section of
this paper, how existing prescriptions treat this divergence, we will
show in section~3 that the divergent series can be treated by Borel
summation, as is the case for threshold
resummation~\cite{frru,afr}. The Borel transform of the series
converges and can be summed. The inversion integral which gives back
the original series diverges, but the divergence can be removed by
including a suitable higher twist term. This leads to a resummed
result of which the original divergent series is an asymptotic
expansion. The ensuing prescription is given in terms of a contour
integral which is easily amenable to numerical implementation. The
result is free of numerical instabilities, and stable upon the
inclusion of subleading corrections. An estimate of the ambiguity on
the resummed results may be obtained from a variation of the
higher-twist term which is included in order to render the results
convergent.  
In section~4 we will compare the
result of our prescription to other existing prescriptions in the case
of the Drell-Yan process, and discuss the ambiguities related to the
resummation procedure. Some results on Fourier transforms are
collected in the Appendix.

\sect{The need for a resummation prescription}

Let us consider a parton--level quantity $\bsigma$ which depends
on a large scale $Q$ and a transverse momentum $\qtvec$, such as
the partonic Drell-Yan differential cross-section $\frac{d\sigma}{d\qt^2}$.
Resummation is necessary because  
the perturbative coefficient of order $n$ in the expansion of $\bsigma$
in powers of $\a(Q^2)$ has the form
\bea
&&\bsigma=\sum_n \a^n(Q^2)\,\bsigma^{(n)}(\qtq,Q^2)
\label{sigmaexp}
\\
&&\quad\bsigma^{(n)}(\qtq,Q^2)
=\left[\frac{P_n(\ln\hqtq)}{\hqtq}\right]_++Q_n(\hqtq)+D_n\delta(\hqtq),
\label{sigmaco}
\eea
where 
\beq
\label{qhatdef}
\hqtq\equiv\frac{\qtq}{Q^2},
\eeq
$P_n(\ln\hqtq)$ is a polynomial of degree
$2n-1$ in $\ln\hqtq$, $Q_n(\hqtq)$ is regular as $\qt\to 0$, and
$D_n$ are constants (see
the Appendix for a definition of the $+$ distribution). Physical
observables are obtained, exploiting collinear factorization, as the
convolution of parton level cross-sections with parton
distributions~\cite{CSS}. When $Q^2$ is large enough, it
sets the scale of parton distributions, and the $\qt$ dependence
is entirely given by the partonic  cross-section. For lower values of
$Q^2$ the scale of parton distributions is set by the impact parameter $b$,
which is Fourier conjugate to $\qt$, the convolution must be performed in
$b$ space, and the Fourier transform must be inverted to obtain physical predictions. 
In either case, the resummation is performed in $b$ space at the level
of partonic observables.

Upon Fourier transformation, $\qtvec$  is replaced by its  Fourier-conjugate
variable, the impact parameter $\vec b$, and the small-$\qt$ region is mapped
onto the large-$b$ region. Large logs of $b$ can then be resummed,
leading to an expression of the form
\beq\label{eq:generic_resumm}
\Sigma\left(\a,\ab L \right)=
\sum_{k=1}^{\infty} h_k(\a)\,(\ab L)^k +O(L^0),
\eeq
where 
\beq
L\equiv\ln\frac{b_0^2}{Q^2 b^2}
\eeq
is the large logarithm which is resummed, and $O(L^0)$ denotes terms
which are not logarithmically enhanced as $b\to\infty$.
For future convenience, we have introduced in the definition of $L$ 
an arbitrary constant $b_0$ (to be discussed below), and we have further
defined 
\beq\label{abdef}
\ab\equiv\beta_0 \aq,
\eeq
$\beta_0$ is the first coefficient of
the QCD beta function,
\bea
&&Q^2\,\frac{\partial\aq}{\partial Q^2}
=-\beta_0\,\a^2(Q^2)\[1+\beta_1\,\a(Q^2)+O(\a^2)\]
\\
&&\beta_0=\frac{33-2n_f}{12\pi},\qquad 
\beta_1=\frac{1}{2\pi}\frac{153-19n_f}{33-2n_f}\,.
\eea
The inverse Fourier transform of $\Sigma$ with respect to $b$ is given by
\beq
\bsigma(\a,\hqtq)=\frac{Q^2}{2\pi}
\int d^2b\,e^{-i\qtvec\cdot\bvec}\,\Sigma(\a,\ab L)
=\int_0^{+\infty}\!d\hat b\,\hat b\,J_0(\hat b\hqt)\,\Sigma(\a,\ab L),
\label{eq:inversefourier}
\eeq
using two-dimensional polar coordinates for $\hat b\equiv b Q$,
and the integral representation of the $0$-th order Bessel function,
\beq
J_0(z)=\frac{1}{2\pi}\int_0^{2\pi}\!d\theta\,e^{-iz\cos\theta}.
\eeq

Now consider specifically the resummation of
\beq
\label{resdist}
\bsigma(\a,\hqtq)= \frac{1}{\hat\sigma_0}\frac{d\hat\sigma}{d\hqtq},
\eeq
where $\frac{d\hat\sigma}{d\hqtq}$ is the partonic
transverse momentum distribution of a massive final state,
and $\hat\sigma_0$ the Born--level total cross-section.
In this case, the $b$-space resummed result has the form~\cite{CSS}
\bea \label{ressig}
&& \Sigma(\a,\ab\,L)=\exp S(\a,\ab\,L),\\
\label{ress}
&&\quad S(\a,\ab\,L)\equiv -
\int_{\frac{b_0^2}{b^2}}^{Q^2}\!\frac{d\mu^2}{\mu^2}\,
\[\ln\frac{Q^2}{\mu^2}\,A(\amu)+B(\amu)\],
\label{exp}
\eea
where 
\beq
A(\a)=A_1\,\a+A_2\,\a^2+\ldots;\qquad
B(\a)=B_1\,\a+\ldots,
\eeq
and the constants $A_i,B_i$ can be determined order by order by
matching to the fixed-order calculation.

The integral in eq.~\eqref{exp} can be performed explicitly, and the
result can then be expanded as
\beq\label{llexp}
S(\a,\ab\,L)=\sum_{i=0}^\infty\ab^{i-1}\,f_i(\ab\,L),
\eeq
where inclusion of the first $k$ orders in the sum corresponds to the
next$^k$-to-leading log (N$^k$LL) approximation.
The LL and NLL
functions $f_0,f_1$ are explicitly given by
\bea
f_0(y)&=&\frac{A_1}{\beta_0}\[\ln(1+y)-y\]
\label{eq:f0}
\\
f_1(y)&=&\frac{A_1\beta_1}{\beta_0^2}
\[\frac{1}{2}\ln^2(1+y)-\frac{y}{1+y}+\frac{\ln(1+y)}{1+y}\]
\nonumber\\
&&-\frac{A_2}{\beta_0^2}\[\ln(1+y)-\frac{y}{1+y}\]
+\frac{B_1}{\beta_0}\,\ln(1+y). \label{eq:f1}
\eea
Note that with $y=\ab L$, using the leading log form of
$\alpha_s(Q^2)$,  
\beq\label{yfromalpha}
1+y=\frac{ \alpha_s(Q^2)}{\alpha_s(b_0^2/b^2)}.
\eeq

It is apparent from eqs.~(\ref{eq:f0},\ref{eq:f1}) that $\Sigma(\a,\ab L)$ 
has a branch cut along the negative real axis
in the complex plane of the variable
$y=\ab L$:
\beq
\label{cut}
{\rm Re\;}(y) \leq -1;\qquad {\rm Im\;}(y)=0.
\eeq
This is due to the fact that the strong coupling blows up when its
argument reaches the Landau pole, so that 
$S(\a,\ab\,L)$
eq.~\eqref{ress} is singular when $b$ becomes large enough, i.e. when
\beq\label{blandau}
b^2\geq\ b_L^2\equiv \frac{b_0^2}{Q^2}\,e^{\frac{1}{\ab}}.
\eeq
At leading order, $b^2_L=\frac{b_0^2}{\Lambda^2}$.
It follows that  the series for
$\Sigma(\a,\ab L)$ eq.~(\ref{eq:generic_resumm}) has a finite radius of
convergence, and the integrand in eq.~(\ref{eq:inversefourier}) is not
analytic in the whole integration range $0\leq\hat b<+\infty$,
so the Fourier inversion integral is not well-defined without a
prescription to treat the singularity.

As mentioned in the introduction, various prescriptions of this
kind have been proposed. Before discussing them, let us show that the
reason why a prescription is needed is the divergence of the expansion
in powers of $\a(Q^2)$ of the
resummed result obtained computing the inverse Fourier transform
eq.~(\ref{eq:inversefourier}) with $\Sigma(\a,\ab\,L)$ eq.~(\ref{ressig}).
To any finite perturbative order, the $\qt$-space resummed result 
is found by expanding eq.~(\ref{ressig}) and inverting the Fourier
transform order by order:
\beq
\bsigma_K(\a,\lb) = \sum_{k=1}^K h_k(\a) \ab^k\,
\frac{Q^2}{2\pi} \int d^2 b \, e^{-i\qtvec\cdot\vec b}\,L^k,
\label{eq:sigmabar}
\eeq
where we have replaced the argument $\hqtq$ of $\bsigma$ by
\beq
\lb\equiv\ln\hqtq=\ln\frac{\qtq}{Q^2}\,.
\eeq

When $K\to\infty$ the series eq.~(\ref{eq:sigmabar})
diverges. To see this, we compute the integrals in
eq.~(\ref{eq:sigmabar}) using eq.~(\ref{ellek})
of the Appendix: 
\bea
&&\bsigma_K(\a,\lb)=\frac{d}{d\hqtq}\,R_K(\a,\lb)
\label{eq:sigmaK}\\
&&\qquad R_K(\a,\lb)=2\,\sum_{k=1}^K h_k(\a) \ab^k 
\sum_{j=0}^{k} \binom{k}{j}\,M^{(j)}(0)\,
\lb^{k-j},
\label{eq:RK}
\eea
where the function $M(\eta)$ is defined in eq.~(\ref{mdef}), 
we have assumed $\hqtq\not=0$, so that distributions can be ignored,
and the term with $j=k$, which leads to a vanishing contribution to
$\bsigma_K(\a,\lb)$,
  has been included in the sum over
$j$ eq.~(\ref{eq:RK}) for later convenience.
We now change the order of summation, and use the identity
\beq
\label{2gammahankel}
\frac{1}{(k-j)!}=\frac{1}{2\pi i}\oint_H d\xi\,e^\xi\,\xi^{-(k-j)-1},
\eeq
where the integration path $H$ is any closed contour which encloses
the origin $\xi=0$. We obtain
\bea
R_K(\a,\lb)&=&2\,\sum_{j=0}^K\frac{M^{(j)}(0)}{j!}
\sum_{k=j}^K \frac{k!}{(k-j)!}\,h_k\,\ab^k\,{\lb}^{k-j}
\\
&=&\frac{1}{\pi i}\oint_H \frac{d\xi}{\xi}\,e^\xi\,
\sum_{j=0}^K\frac{M^{(j)}(0)}{j!}\,\(\frac{\xi}{{\lb}}\)^j
\sum_{k=j}^K k!\,h_k\,\(\frac{\ab\lb}{\xi}\)^k\,.
\label{eq:Rdiv}
\eea

Because of the singularity eq.~(\ref{cut}), the
power series eq.~\eqref{eq:generic_resumm} has a finite radius of
convergence equal to one
\beq
\lim_{k\to\infty}\abs{\frac{h_{k+1}}{h_k}}=1,
\eeq
which immediately implies the vanishing of the radius of convergence
of the sum over  $k$ in eq.~\eqref{eq:Rdiv}.

The situation is thus similar to that which is encountered in
threshold resummation~\cite{cmnt,frru,afr}: the resummation is performed on
quantities which are related by Mellin transformation to the physical
ones, but the resummed results cannot be expressed as a  Mellin
transform of some function. Namely,  their inverse Mellin
transform does not exist, as
a consequence of the fact that the inverse Mellin transform of their
expansion in powers of $\a(Q^2)$ diverges. 
In the present case, the divergence of the perturbative expansion
implies that the Fourier inversion integral is ill-defined;
of course the problem disappears if one retains only a finite
number of terms in the resummed expansion~\cite{Kulesza:1999sg,Kulesza:1999gm}.
Various commonly used prescriptions replace the ill-defined integral
with a well defined one, as we now review. In the next section, we
construct a prescription which is instead based on the idea of
replacing the divergent series with a convergent one through the Borel
summation method. In the last section we will compare the various
prescriptions and in particular the way they treat the divergence of
the perturbative series.

In the prescription of ref.~\cite{CSS},
the variable $b$ is replaced by a function $b_\star(b)$ which approaches
a finite limit $b_{\rm lim}\leq b_L$ as $b\to\infty$, such as for example
\beq
\label{eq:bstar}
b_\star=\frac{b}{\sqrt{1+(b/b_{\rm lim})^2}}.
\eeq
In this way, the cut eq.~(\ref{cut}) is never reached. This procedure
has some degree of arbitrariness in the choice of the function $b_\star(b)$,
which is interpreted as a parametrization of non-perturbative effects,
whose size can be estimated by varying $b_\star$, for
 instance by changing the value of $b_{\rm lim}$.
The matching of this prescription  
to the fixed-order result is however numerically unstable, as pointed out in
ref.~\cite{EV}.

A different possibility~\cite{EV} is based on the observation that
if only the leading log contribution (i.e. the terms with $j=0$) are
included in eq.~(\ref{eq:RK}), then the series converges, and its sum
can in fact be computed in closed form, with the result (see
eq.~(\ref{llft}) of the Appendix)
\beq
\bsigma_{\rm LL}(\a,\lb)=2 \frac{d}{d\hqtq}\Sigma\(\a,\ab\lb\).
\label{eq:EV}
\eeq
Equation~\eqref{ress} implies that $S (\a,\ab\,L)$ depends on $b^2$
through $\a(1/b^2)$. Therefore, using eq.~(\ref{yfromalpha}), 
the LL expression eq.~\eqref{eq:EV} is seen to become a function of
$\a(\qtq)$. Therefore, the leading log truncation of the perturbative
expansion   in powers of
$\alpha_s(Q^2)$ eq.~(\ref{eq:Rdiv}) has 
a  finite radius of convergence, set by the  Landau pole
\beq
\qtq>Q^2\exp\(-\frac{1}{\ab}\)=\Lambda^2,
\label{eq:qtrange}
\eeq
where the last equality holds at leading order. 

The main defect of this result is that it is subject to large
next-to-leading log corrections. In fact, the NLL Fourier inversion
integral
can also be computed
in closed form~\cite{FNR}. The result (given in eq.~(\ref{nllft})) differs
sizably from the LL result even for relatively large values of $\qt$
(several GeV for $Q=100$~GeV), as
we shall see explicitly in Sect.~4 below. 
In fact, it turns out that the NLL correction diverges at a value of
$\qt$ which is an increasing function of the scale $Q$. This
instability can be 
understood as a consequence of the fact that the truncation of the 
resummed result to  finite 
logarithmic accuracy leads to an expansion in powers of $\a(\qtq)$
with coefficients depending on $\ln(\qt/Q)$, where higher powers of
$\a(\qtq)$ correspond to higher logarithmic orders. Such an expansion is necessarily
poorly  behaved at low $\qt$,
all the more so when the scale ratio $\qt/Q$ is large.
Performing the Fourier inversion to leading or next-to-leading 
logarithmic accuracy thus removes
the divergence of the series eq.~\eqref{eq:sigmabar}: this is
analogous to what is found in the case of threshold resummation, where
it can be shown~\cite{frru} that the divergence of resummed results is
removed if the Mellin inversion is performed to any finite logarithmic
accuracy. However, the ensuing results are then perturbatively unstable.

A yet different way of treating the divergence has been proposed more
recently in ref.~\cite{MP}, along the lines of the so--called Minimal
Prescription of threshold resummation~\cite{cmnt}.  The basic idea
here is that to any finite perturbative order, when the divergent
series is replaced by a finite sum, one may choose the integration
path in such a way that it avoids the singularities which appear at
the resummed level. The result of the Fourier (or respectively Mellin)
inversion is then unchanged to any finite perturbative order, but it
becomes finite at the resummed level. It can be further
shown~\cite{cmnt} that the divergent perturbative expansion of the
resummed expression is asymptotic to the result obtained in this way.
This prescription is widely used~\cite{higgs}: whereas in the case of
threshold resummation it leads to dependence of resummed physical
results on a kinematically unaccessible region (albeit by
power--suppressed terms), in the case of transverse momentum
resummation its only shortcoming is speed limitation in its numerical
implementation.

\sect{The Borel prescription}

We now turn to the construction of a prescription which extends 
to transverse momentum resummation the Borel prescription
proposed in refs.~\cite{frru,afr} for the
resummation of threshold logarithms. The
basic idea is to tackle directly the divergence of the
series~(\ref{eq:RK},\ref{eq:Rdiv}) by summing it through the Borel method. 

To do this,  we take the Borel transform 
of eq.~(\ref{eq:Rdiv}) with respect to~$\ab$. This amounts
to the replacement $\ab^k\to w^{k-1}/(k-1)!$, where $w$ is the Borel
variable conjugate to $\ab$. We obtain
\beq
\hat R_K(w,\lb)=\frac{1}{\pi i}\oint_H\frac{d\xi}{\xi^2} \,e^{\lb\xi}\,
\sum_{j=0}^K\frac{M^{(j)}(0)}{j!}\, \xi^j 
\sum_{k=1}^K k\,h_k\, \left(\frac{w}{\xi}\right)^{k-1} ,
\label{eq:ultima}
\eeq
where in comparison to eq.~(\ref{eq:Rdiv}) we have rescaled the
integration variable  $\xi\to\lb\xi$, and we have included all
terms with $1\leq k\leq j-1$, which vanish upon contour
integration.

\begin{figure}
\centering
\epsfig{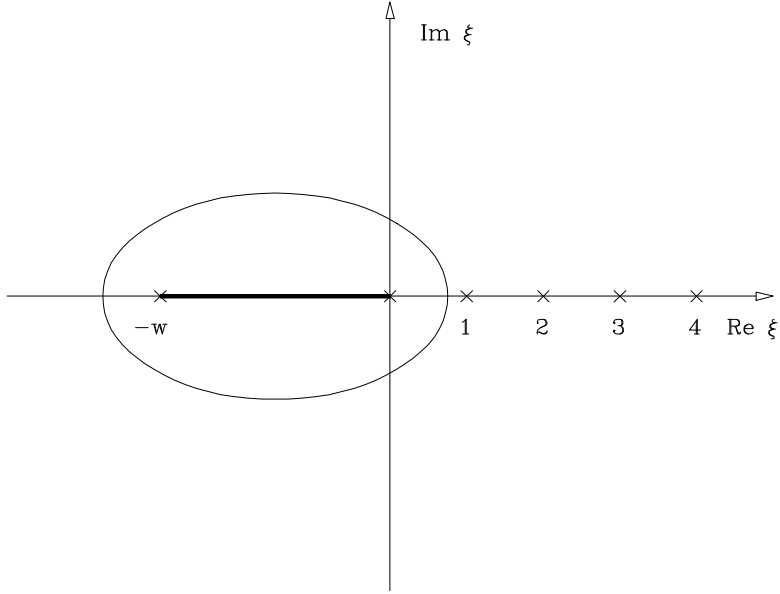}  
\caption{The integration contour $H$ in eq.~(\ref{eq:tappeto}). \label{contour}}
\end{figure}

Both sums in eq.~\eqref{eq:ultima} are convergent as $K\to\infty$. Indeed,
\begin{align}
&\sum_{k=1}^\infty k\,h_k\,\left(\frac{w}{\xi}\right)^{k-1}
=\xi\,\frac{d}{dw}\,\Sigma\left(\a,\frac{w}{\xi}\right)
&&\text{for}\;\; \abs{\frac{w}{\xi}}<1
\\
&\sum_{j=0}^\infty\frac{M^{(j)}(0)}{j!}\, \xi^j =M(\xi)
&&\text{for}\;\; \abs{\xi}<1,
\end{align}
the last condition being due to the simple pole of $M(\xi)$ at $\xi=1$.
Thus,
\beq
\hat R(w,\lb)=\lim_{K\to\infty}\hat R_K(w,\lb)
=\frac{1}{\pi i}
\oint_H \frac{d\xi}{\xi}\,e^{\lb\xi}\,M(\xi)\,
\frac{d}{dw}\,\Sigma\left(\a,\frac{w}{\xi}\right),
\label{eq:tappeto}
\eeq
provided the contour $H$ is chosen so that
\beq
w<\abs{\xi}< 1 .
\eeq
Since $M(\xi)$ has no singularities on the negative real
axis, and $\Sigma(\a,w/\xi)$ has a branch cut on the real $\xi$ axis
between $-w$ and $1$, the integration contour can now be deformed
so that $\hat R(w,\lb)$ is well defined for all positive values of $w$
(see fig.~\ref{contour}).

The original function eq.~(\ref{eq:RK}) is recovered by inverting
the Borel transform:
\beq\label{eq:inversion}
R(\a,\lb) = \int_0^\infty dw \, e^{-\frac{w}{\ab}}\,\hat R(w,\lb) .
\eeq
The inversion integral is divergent at $w\to\infty$.
This is easily seen by inspection of fig.~\ref{contour}: as $w$
becomes large, the branch cut extends to the left, and the integration
contour is pushed towards large negative values of $\xi$, where
$M(\xi)$ oscillates with a factorially growing amplitude.

We regulate the integral by cutting it off at $w=C$. We thus get 
\beq
R^C(\a,\lb)=\frac{1}{\pi i}\oint_H \frac{d\xi}{\xi}
\,M(\xi)\,e^{\lb\xi}\,\int_0^C dw\,e^{-\frac{w}{\ab}}\,
\frac{d}{dw}\Sigma\(\a,\frac{w}{\xi}\),
\label{eq:RB}
\eeq
which is the Borel prescription for transverse momentum
resummation.
The result can be equivalently rewritten by doing a partial integration as
\beq
R^C(\a,\lb)=\frac{1}{\pi i}\oint_H \frac{d\xi}{\xi}
\,M(\xi)\,e^{\lb\xi}\,\left[
e^{-\frac{C}{\ab}}\Sigma\left(\a,\frac{C}{\xi}\right)
+\frac{1}{\ab}\int_0^C dw \,
e^{-\frac{w}{\ab}}\,\Sigma\left(\a,\frac{w}{\xi}\right)
\right],
\label{eq:RBpartial}
\eeq
which may be more convenient for numerical implementations in that it
depends directly on the physical observable  $\Sigma$, rather than its
derivative.
Equation~(\ref{eq:RBpartial}), and its equivalent form eq.~(\ref{eq:RB}),
are the main result of this paper. It is interesting to observe that
if we integrate by parts before cutting off the integral, then the
surface term vanish. We then end up with the alternative resummation
\beq
R^{C^\prime}(\a,\lb)=\frac{1}{\pi i}\oint_H \frac{d\xi}{\xi}
\,M(\xi)\,e^{\lb\xi}\frac{1}{\ab}\int_0^C \!dw\, 
e^{-\frac{w}{\ab}}\,\Sigma\left(\a,\frac{w}{\xi}\right).
\label{eq:RBpartialprime}
\eeq
As we shall see shortly, this is an equally valid prescription.

In order to see that this is a valid resummation prescription,
consider the truncation to order $K$ of eq.~(\ref{eq:inversion}), namely
\bea
R^C_K(\a,\lb)&\equiv&\int_0^C dw\,e^{-\frac{w}{\ab}} \hat R_K(w,\lb) 
\nonumber\\
&=&2\,\sum_{j=0}^K\frac{M^{(j)}(0)}{j!}
\sum_{k=j}^K \frac{k!}{(k-j)!}\,h_k\,\ab^k\,
\frac{\gamma\(k,\frac{C}{\ab}\)}{(k-1)!} \,\lb^{k-j},
\label{eq:Rht}
\eea
where
\beq
\gamma(k,z)=\int_0^z dw\, e^{-w} w^{k-1}=(k-1)!
\(1- e^{-z} \sum_{n=0}^{k-1}\frac{z^n}{n!} \)
\eeq
is the truncated gamma function.
The difference between the original $R_K(\a,\lb)$
eq.~\eqref{eq:sigmaK} and its  Borel resummation $R^C_K(\a,\lb)$ is
\bea
R_K^{\rm ht}(\a,\lb;C) &\equiv& R_K(\a,\lb) - R^C_K(\a,\lb)
\nonumber\\
&=&2\,e^{-\frac{C}{\ab}}
\sum_{j=0}^K\frac{M^{(j)}(0)}{j!}
\sum_{k=j}^K \frac{k!}{(k-j)!}\,h_k\,\ab^k\,\lb^{k-j} \,
\sum_{n=0}^{k-1}\frac{1}{n!}\(\frac{C}{\ab}\)^n.
\label{eq:ht}
\eea

Because
\beq\label{twist}
e^{-\frac{C}{\ab}}=
\( \frac{\Lambda^2}{Q^2} \)^C \left[ 1+ O\(\aq\) \right],
\eeq
$R_K^{\rm ht}(\a,\lb;C)$ is seen to be power--suppressed at large
$Q^2$ (higher twist): cutting off the $w$ integration at $w=C$
is equivalent to the inclusion of a higher twist term, which cancels
the divergence of the resummed expression. Specifically,
$R_K^{\rm ht}(\a,\lb;C)$ is a twist-$t$ contribution with
\beq\label{tfromc}
t=2\,(1+C),
\eeq
the  choice $C=1$ corresponds to the inclusion of a twist-four
term.  Moreover, it is apparent from  eq.~\eqref{eq:ht} that
\beq\label{htasymp} 
R_K^{\rm ht}(\a,\lb;C) \twiddles{\a\to 0}e^{-\frac{C}{\ab}},
\eeq
which vanishes faster than any power of $\a$ as $\a\to0$. It follows that
the original divergent
$R_K(\a,\lb)$ is an asymptotic expansion of the Borel-resummed result
 $R^C(\a,\lb)$ eqs.~(\ref{eq:RBpartial},\ref{eq:RB}).

Furthermore, the alternative prescription $R^{C^\prime}(\a,\lb)$
eq.~(\ref{eq:RBpartialprime}) differs from $R^{C}(\a,\lb)$
eq.~(\ref{eq:RBpartial}) by the first term in square brackets
in~(\ref{eq:RBpartial}), which is a finite higher-twist
contribution. Hence, the two prescriptions correspond to two
inequivalent but equally acceptable
regularizations of the divergent sum which differ by finite terms, and
are both asymptotic sums of the divergent series.

The main features of the Borel prescription can be appreciated by
considering as an explicit example of a resummed quantity
$\Sigma(\a,\ab L)=\gamma_{\rm LL}(\a,\ab L)$, with
\beq
\gamma_{\rm LL}(\a,\ab L)\equiv \frac{dS_{\rm LL}(\a,\ab L)}{d\ln Q^2},
\eeq
and $S_{\rm LL}(\a,\ab L)$ given by eqs.~(\ref{ress},\ref{llexp})
evaluated at the leading log level~(\ref{eq:f0}), namely
\beq
\gamma_{\rm LL}(\a,\ab L)=\frac{A_1}{\beta_0}\,\ln(1+\ab\,L).
\eeq
Substituting this form of $\Sigma(\a,\ab L)$ in eq.~(\ref{eq:RB}), the
associate $\qt$-space physical observable computed with the Borel
prescription is found to be 
\beq
\bar\gamma^C_{\rm LL}(\a,\lb)
=\frac{A_1}{\beta_0}\,\frac{1}{\hqtq}
\int_0^{C} dw\,e^{-\frac{w}{\ab}}
\frac{1}{\pi i}\oint_H d\xi
M(\xi)\,e^{\lb\xi}\,\frac{1}{\xi+w}.
\eeq
The $\xi$ integral is easy to calculate, because the integrand
has only a simple pole at $\xi=-w$:
\beq  \bar\gamma^C_{\rm LL}(\a,\lb)
=\frac{2A_1}{\beta_0}\,\frac{1}{\hqtq}\,
\int_0^{C} dw\,\(\frac{\Lambda^2}{\qtq}\)^w\,M(-w),
\eeq
where we have used the leading-log expression of the running coupling.
It is thus clear that the divergent integration is cut off by the
inclusion of a power--suppressed contribution
\bea
\bar\gamma^{\rm ht}_{\rm LL}(\a,\lb;C)
&=&\frac{2A_1}{\beta_0}\,\frac{1}{\hqtq}\,
\int_C^{+\infty} dw\,\(\frac{\Lambda^2}{\qtq}\)^w\,M(-w)\nonumber\label{qthta}\\
&=&\frac{2A_1}{\beta_0}\,\frac{1}{\hqtq}\,\(\frac{\Lambda^2}{\qtq}\)^C\,
\int_0^{+\infty} dw\,\(\frac{\Lambda^2}{\qtq}\)^w\,M(-w-C).
\label{qtht}\eea
Note that the suppression is by powers of  $\frac{\Lambda^2}{\qt^2}$:
at finite order $K$ 
the higher twist contribution is suppressed
by a power of $\frac{\Lambda^2}{Q^2}$, as shown in eq.~(\ref{eq:ht}),
but when resummed to all
orders, the scale $Q^2$ is replaced by an effective scale $\qt^2$.

\sect{Comparison of resummation prescriptions}

Let us now compare the results found using the Borel prescription to
those of other prescriptions, with the dual goal    of understanding
the advantages and disadvantages of various methods, and of assessing 
the ambiguity which is intrinsic to the resummation of  a divergent
expansion.

\begin{figure}
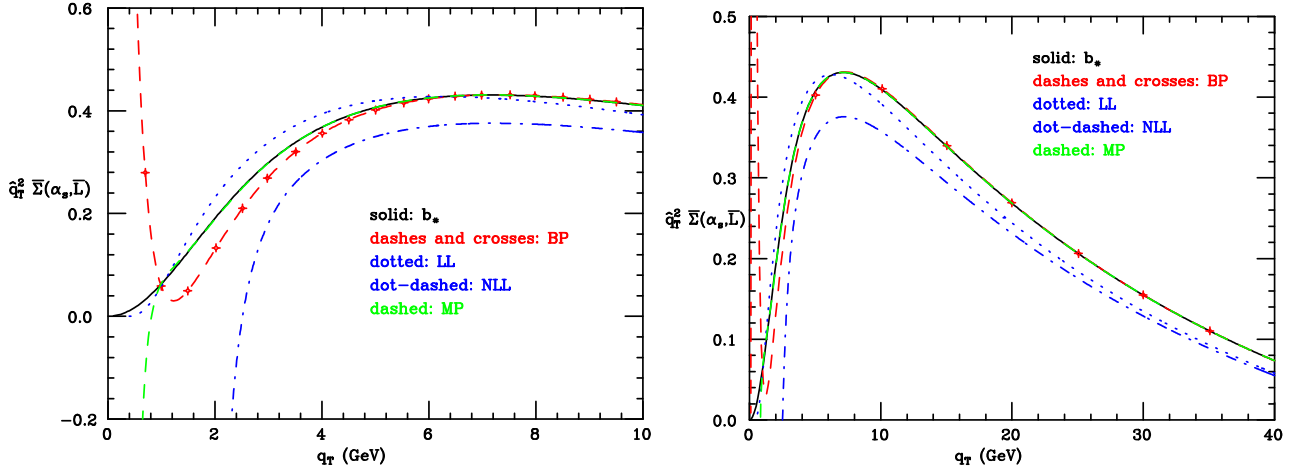

\centering
\includegraphics[width=.5\linewidth]{sigma-zoom.ps}
\includegraphics[width=.48\linewidth]{sigma.ps}
\caption{The NLL partonic resummed Drell-Yan transverse momentum
  distribution  computed with various resummation prescriptions with
  $Q^2=10^4$~GeV$^2$ and in a narrow (left) and wide (right) range of
$\qt$.\label{fig:sigma}}
\end{figure}
First, we look at a typical resummed observable. Namely, we consider
the transverse momentum distribution of Drell-Yan
pairs, eq.~\eqref{resdist}, 
which we evaluate at the partonic resummed next-to-leading log
level, i.e. using eq.~(\ref{ressig}) with $S(\a,\ab\,L)$ computed
including the first two terms in eq.~(\ref{llexp}), 
given in eqs.~(\ref{eq:f0},\ref{eq:f1}) with~\cite{DS,acoeffs}
\bea
\label{a1val}
&&A_1=\frac{C_F}{\pi}
\\ \label{a2val}
&&A_2=\frac{1}{\pi^2}\left( \frac{67}{9}-\frac{\pi^2}{3}-\frac{10}{27} n_f
   +\frac{8\pi}{3}\,\beta_0\,\ln\frac{b_0 e^{\gamma_E}}{2} \right) 
\\ \label{w1val}
&&B_1=\frac{2C_F}{\pi}\, \ln\frac{b_0 e^{\gamma_E-3/4}}{2  }\,.
\eea

The results are displayed in fig.~\ref{fig:sigma}, for $Q=100$~GeV.
The two lower curves at large $\qt$ in this figure
correspond to those found using respectively eqs.~(\ref{llft}) and
eqs.~(\ref{nllft}) of the Appendix, namely, to inverting the Fourier
transform to leading and next-to-leading log accuracy (with $b_0=2
e^{-\gamma_E}$). The  sizable difference between these two results even for $\qt$
as large as 10~GeV shows the instability of the truncation of the
Fourier transform to finite log accuracy discussed in the introduction
and first stressed in ref.~\cite{FNR}.  

The other prescriptions displayed in fig.~\ref{fig:sigma} are the $b_\star$
prescription, where the Fourier inversion is performed after replacing $b$ with
$b_\star$ eq.~(\ref{eq:bstar}), with $b_{\rm lim}=b_L$, where
$b_L=7.2$~GeV$^{-1}$
is the NLO Landau pole eq.~(\ref{blandau}); the minimal
prescription (MP) where the Fourier inversion is performed along the
deformed path of ref.~\cite{MP}, and the Borel prescription
eq.~(\ref{eq:RB}) with $C=1$.

\begin{figure}
\centering
\epsfig{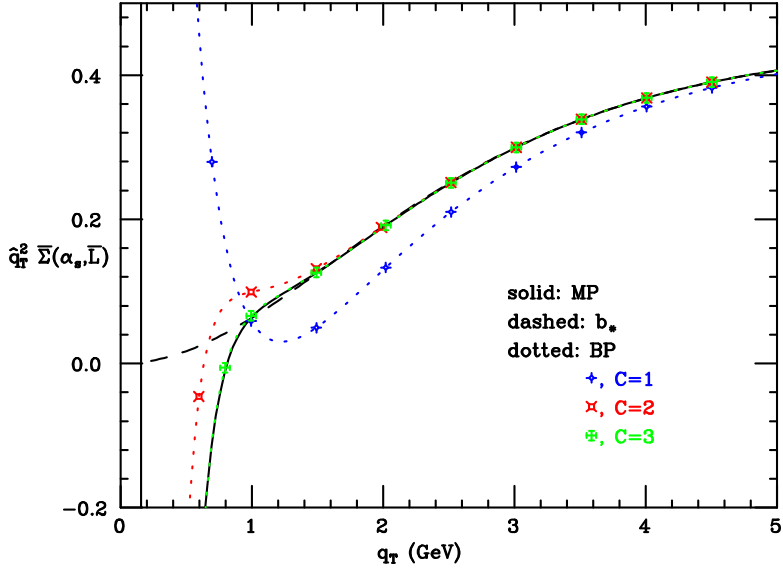}  
\caption{Dependence of the results shown in fig.~\ref{fig:sigma}
for the Borel prescription on the parameter~$C$. The vertical line at 
$\qt=156$~MeV
indicates the position of the Landau pole.\label{fig:Cdep}}
\end{figure}
In fig.~\ref{fig:Cdep} we further show the dependence of the Borel
prescription on the parameter $C$ which characterizes the higher twist
term included in the resummation 
eqs.~(\ref{twist},\ref{tfromc}), as it is varied  between twist four and twist
eight. Because all these choices provide valid resummation prescriptions,
this variation provides an estimate of the ambiguity which is intrinsic of the
resummation procedure: indeed, the $b_\star$ and minimal prescription,
also shown in this figure, are well within the band of variation as
$\qt\to0$. 
These plots show  that the ambiguity in the resummation procedure
is negligible for
$\qt\gtrsim 5$~GeV, it remains small for $\qt\gtrsim 2$~GeV, and it only
blows up as $\qt$ approaches the Landau pole.

We can further elucidate the origin of these results by studying the
effect of the various prescriptions when the divergent sum
eq.~(\ref{eq:sigmabar}) is truncated, so the Fourier inversion can be
performed term by term. Consider specifically the first term in the
series, namely, the
inverse Fourier transform of $L$. The exact result is given by eq.~\eqref{ellek}
for $k=1$,
\beq
\frac{1}{2\pi}\int d^2\hat b\,e^{-i\hqt\cdot \hat b}\,
\ln\frac{b_0^2}{\hat b^2}=\frac{2}{\hqtq}.
\eeq
The MP reproduces this  exact result, because
$\ln(b_0^2/\hat b^2)$ is analytic on the positive real
$\hat b$ axis, and a deformation of the integration contour has no
effect; a branch cut on the positive real $\hat b$ axis
only arises after summation of the whole series.

The Borel prescription yields instead
\beq\label{borellog}
\[\frac{1}{2\pi}\int d^2\hat b\,e^{-i\hqt\cdot \hat b}\,
\ln\frac{b_0^2}{\hat b^2}\]_{\rm BP}
=\frac{2}{\hqtq}\,\(1-e^{-\frac{C}{\ab}}\)
\eeq
as one can see by setting $h_1=\frac{1}{\ab}$ and $h_k=0$ for all
$k\not=1$ in eq.~\eqref{eq:Rht}. The exact result is modified by
the introduction of a correction of twist $2(1+C)$. Note that the
higher twist correction is tiny at large $Q^2$, of order $10^{-6}$ for
$C=1$ and
$Q^2=10^4$~GeV$^2$. If we use the alternative Borel prescription $R^{C^\prime}(\a,\lb)$
eq.~(\ref{eq:RBpartialprime}) we get instead 
\beq\label{borellogprime}
\[\frac{1}{2\pi}\int d^2\hat b\,e^{-i\hqt\cdot \hat b}\,
\ln\frac{b_0^2}{\hat b^2}\]_{\rm BP^{\prime}}
=\frac{2}{\hqtq}\,\[1-e^{-\frac{C}{\ab}}
\left(1+\frac{C}{\ab}\right)\]\,.
\eeq
so the two prescriptions are indeed seen to differ by a higher twist term.

Finally, the result of
the replacement of $b$ by $b_\star$ eq.~\eqref{eq:bstar}
can be computed analytically
in terms of the Bessel function $K_1$:
\bea
\frac{1}{2\pi}\int d^2\hat b\,e^{-i\hqt\cdot \hat b}\,
\ln\frac{b_0^2}{\hat b_\star^2}
&=&
\frac{1}{2\pi}\int d^2\hat b\,e^{-i\hqt\cdot \hat b}\,
\ln\[\frac{b_0^2}{\hat b^2}\(1+\frac{\hat b^2}{\hat b_{\rm lim}^2}\)\]
\nonumber\\
&=&\frac{2}{\hqtq}\,\[1-\hat b_{\rm lim}\hqt\,K_1(\hat b_{\rm lim}\hqt)\].
\label{eq:bstarlog}
\eea
Using the asymptotic behaviour $K_1(z)\twiddles{z\to\infty}e^{-z}/\sqrt{z}$,
we see that the correction factor in eq.~\eqref{eq:bstarlog} vanishes
faster than any power of $1/(b_{\rm lim} \qt)$ for $\qt\gg 1/b_{\rm lim}$.

For higher order powers of $L$ the same qualitative behaviour is found
using the various prescriptions discussed here. Namely, the MP gives
the exact Fourier
transform eq.~(\ref{ellek}); the BP gives a result which differs from
it by a higher twist term, and the $b_\star$ prescription gives a
result which differs from it by a term which is exponentially
suppressed in $1/(b_{\rm lim}\qt)$. 

We thus see that the way different prescriptions tackle the divergence
of the perturbative expansion is the following. In the LL and NLL
case, the divergent series eq.~(\ref{eq:RK}) is made convergent by
truncating the Fourier inversion to finite order, i.e. by only
retaining a finite number of terms in the inner sum over $j$. This, as
discussed in Section~2, leads effectively to an expansion in powers of
$\a(\qtq)$ which has very poor convergence properties at small $\qt$
even when $Q$ is large. The MP and BP both provide an asymptotic sum
of the divergent series: the BP removes the divergence by inclusion of
a higher twist term, and the MP by a suitable analytic
continuation, which corresponds~\cite{cmnt} to the inclusion of terms
which are more suppressed than any power of $Q^2$.  
At large $Q^2$, the higher twist term of the BP is
negligible so these two prescriptions are essentially
indistinguishable when  applied to convergent series. When applied to
the divergent resummed expansion displayed in
figs.~\ref{fig:sigma}-\ref{fig:Cdep} they only
differ in the region where $\qt$ approaches the Landau pole, so the
high--order behaviour of the series become relevant. Finally, the
$b_\star$ prescription modifies the divergent series by inclusion of a
term which is more suppressed than any power of $1/(b_{\rm lim}\qt)$.
When applied
to a convergent series, this prescription produces a result that
differs sizably from that of the BP when $\qtq\ll Q^2$ and it approaches the
Landau pole: this is because the scale of the correction term is
set by $Q^2$ for the BP, and by $\qt^2$ for the $b_\star$ prescription.
At the resummed level, however, the effective
scale of power suppressed terms becomes $\qt^2$ also for the BP
(compare eq.~(\ref{qtht})), so all resummation prescriptions lead
essentially to the same result.

\section{Summary}

We have constructed a resummation prescription for transverse momentum
distributions which extends to this case the Borel prescription
previously proposed for threshold
resummation~\cite{frru,afr}. The construction is based on the
observation that the reason why a resummation prescription is needed in
the first place is that the perturbative expansion of resummed results
in $\qt$ space in powers of $\a(Q^2)$ diverges. The Borel prescription
tackles this divergence by summing the convergent Borel transform of
the divergent series, and then making the Borel inversion finite by
inclusion of a higher twist term. The original divergent series is an
asymptotic expansion of the result obtained thus. 
The Borel prescription is easily amenable to
numerical implementation; being based on a $b$-space resummation it is
easy to match to fixed--order results, and it is perturbatively
stable. 

There is some freedom in this prescription, parametrized by a real
parameter $C$, related to the twist $t$ of the term included in order
to obtain convergence by $t=2(C+1)$. Whereas $C$ may be chosen to take
any value, it is convenient to choose a value which corresponds to
twists which already appear in the expansion of the observable
being considered. Indeed, physical observables must be independent
of the choice of $C$, and thus if an unphysical twist term is
introduced, it must be compensated by an equal and opposite power
suppressed term which is thereby artificially introduced by this
choice.

Comparison of the Borel prescription to other available resummations,
such as the minimal prescription or the $b_\star$ method, shows that at
large $Q^2$ they lead to results which are extremely stable and which
only differ when $\qt$ approaches the Landau pole. In fact, variation
of the parameter $C$ of the Borel prescription provides a
reliable estimate of the ambiguity in the resummation procedure. For
$\qt\gtrsim$~2~GeV this ambiguity appears to be negligibly small, even
in the region of a few GeV where the impact of the resummation is
sizable. This is in contrast to the case of threshold resummation,
where it was found~\cite{afr} that the ambiguity is almost as large as
the effect of the resummation itself in most of the kinematic region
where the resummation is relevant. 

Our results contradict the widespread
prejudice that transverse momentum resummation is affected by sizable
ambiguities, and it shows that, at least as long as $Q$ is as large
as the $W$ mass  and
$\qt$ as large as the nucleon mass perturbative resummation of
transverse momentum distributions provides reliable and stable
results. The Borel prescription provides a new method for performing
this resummation which has more stable matching properties than the
$b_\star$ prescription and might be numerically advantageous over the
widely used minimal prescription.
\bigskip

{\bf Acknowledgements:} We thank G.~Altarelli for discussions.
This work was partly supported by the European network HEPTOOLS under contract
MRTN-CT-2006-035505 and by a PRIN2006 grant (Italy).

\eject
\appendix
\sect{Appendix}
In this appendix, we collect some results on two-dimensional
Fourier transforms of powers of logarithms.

First, we compute the exact Fourier transform with respect to $\hat b$ 
of the $k$--th power of
$\ln^k\frac{b_0^2}{\hat b^2}$ (with $b_0$ a constant). We get 
\beq
\frac{1}{2\pi}\int d^2\hat b\,e^{-i\hqt\cdot \hat b}\,
\ln^k\frac{b_0^2}{\hat b^2}
=2\,\,M^{(k)}(0)\,\delta(\hqtq)
+2\,\sum_{j=0}^{k-1}\binom{k}{j}\,M^{(j)}(0)\,
\left[\frac{d}{d\hqtq}\ln^{k-j}\hqtq\right]_+,
\label{ellek}
\eeq
where
\beq\label{mdef}
M(\eta)=\(\frac{b_0^2}{4}\)^\eta\frac{\Gamma(1-\eta)}{\Gamma(1+\eta)},
\eeq
and the $+$ distributions are defined by
\beq
\int_0^1 d\hqtq\,\left[D\(\hqtq\)\right]_+=0.
\label{plusdefgen}
\eeq

In order to prove eq.~(\ref{ellek}), we define a generating function
\beq
\chi(\hat b,\eta)=\left(\frac{b_0^2}{\hat b^2}\right)^\eta;
\qquad
L^k=\ln^k\frac{b_0^2}{\hat b^2}=
\left.\frac{\partial^k}{\partial\eta^k}\,\chi(\hat b,\eta)\right|_{\eta=0}.
\eeq
We have
\beq
\frac{1}{2\pi}\int d^2\hat b\,e^{-i\hqt\cdot\hat b}\,\chi(\hat b,\eta)
=\int_0^{+\infty}d\hat b\,\hat b\,J_0(\hat b \hqt)\,
\left(\frac{b_0^2}{\hat b^2}\right)^\eta,
\eeq
where we have used polar coordinates for $\hat b$, and the integral
representation of the 0-th order Bessel function 
\beq
J_0(z)=\frac{1}{2\pi}\int_0^{2\pi}d\theta\,e^{-iz\cos\theta}.
\eeq
The integral can be computed by means of the identity
\beq
\int_0^{+\infty}dx\,x^\mu\,J_\nu(ax)
=2^\mu\,a^{-\mu-1}\,
\frac{\Gamma\left(\frac{1}{2}+\frac{\nu}{2}+\frac{\mu}{2}\right)}
{\Gamma\left(\frac{1}{2}+\frac{\nu}{2}-\frac{\mu}{2}\right)}
\label{GR}
\qquad
a>0;\quad -{\rm Re}\,\nu-1<{\rm Re}\,\mu<\frac{1}{2}.
\eeq
We find
\beq
\frac{1}{2\pi}\int d^2\hat b\,e^{-i\hqt\cdot\hat b}\,\chi(\hat b,\eta)=
2\,\eta\,M(\eta)\,\(\hqtq\)^{\eta-1}.
\label{eq:invchi}
\eeq
We may now replace
\beq
\(\hqtq\)^{\eta-1}=
\left[\(\hqtq\)^{\eta-1}\right]_++\frac{1}{\eta}\,\delta(\hqtq),
\eeq
consistent with the definition eq.~(\ref{plusdefgen}). We get
\beq
\frac{1}{2\pi}\int d^2\hat b\,e^{-i\hqt\cdot\hat b}\,\chi(\hat b,\eta)
=2\,M(\eta)\,\left\{\delta(\hqtq)
+\left[\frac{d}{d\hqtq}\(\hqtq\)^\eta\right]_+\right\}.
\eeq
Evaluating the $k$-th derivative of both sides
with respect to $\eta$ at $\eta=0$ leads immediately to
eq.~(\ref{ellek}). 
Note that the term $j=k$ 
is excluded from the sum  because it vanishes upon differentiation with
respect to $\hqtq$. For $\hqtq$ strictly larger than zero, both the
term proportional to $\delta(\hqtq)$ and the $+$ prescription have no effect.

Let us now turn to the evaluation of the Fourier transform to fixed
logarithmic accuracy.
Equation~(\ref{ellek}) shows that the Fourier transform of the $k$-th
power of $\ln b$ is proportional to $1/\hqtq$ times the $(k-1)$-th power of 
the log of the Fourier
conjugate variable $\ln \hqtq$ (leading log approximation), 
but also includes terms proportional to all lower
powers of this log.  The  N$^n$LL approximation corresponds to
including terms up to $j=n$ in the sum in eq.~(\ref{ellek}),
i.e. such that the power of $\ln \hqtq$ is by $n+1$ units lower than
the power of $\ln(b_0^2/\hat b^2)$. 

The NLL and N$^2$LL approximations are particularly simple due to the
fact that
\bea \label{nllvan}
&&M^{(1)}(0)=\ln\frac{b_0^2}{4}+2\,\gamma_E
\\
\label{nnllvan}
&&M^{(2)}(0)=\(\ln\frac{b_0^2}{4}+2\,\gamma_E\)^2
\eea
where $\gamma_E\approx0.5772$ is the Euler constant.
It follows in particular that if 
$b_0=2\,e^{-\gamma_E}$, the NLL and NNLL terms in
eq.~(\ref{ellek}) vanish~\cite{DS}.

A useful form of the N$^n$LL approximation can be obtained noting that
\beq
M^{(j)}(0)=\int_0^\infty dx\,J_1(x)\,\ln^j\frac{b_0^2}{x^2}.
\eeq
It follows that eq.~(\ref{ellek}) (for $\hqtq>0$, i.e. neglecting
distributions) can be rewritten as 
\beq
\frac{1}{2\pi}\int d^2\hat b\,e^{-i\hqt\cdot\hat b}\,
\ln^k\frac{b_0^2}{\hat b^2}
=2\,\frac{d}{d\hqtq}\int_0^\infty dx\,J_1(x)\,
\(\ln\hqtq+\ln\frac{b_0^2}{x^2}\)^k.
\label{result2}
\eeq
The N$^n$LL approximation can then be obtained by retaining the first
$n$ terms in the binomial expansion  of
$\(\ln\hqtq+\ln\frac{b_0^2}{x^2}\)^k$ in this equation.

This result is particularly useful in that it allows the computation in
closed form of
some  Fourier transforms of  generic functions to fixed logarithmic
accuracy.
Specifically, consider a function 
\beq\label{gfromf}
F(L)=\sum_{k=0}^\infty F_k\,L^k.
\eeq
Its Fourier transform to LL accuracy is given by
\beq\label{llft}
\left[\frac{1}{2\pi}\int d^2\hat b\,e^{-i\hqt\cdot\hat b}\,F(L)\right]_{\rm LL}
=2\,\frac{d}{d\hqtq}\sum_{k=0}^\infty F_k\,\int_0^\infty dx\,J_1(x)\,\ln^k\hqtq
=2\,\frac{d}{d\hqtq}F(\ln\hqtq).
\eeq
This result was given in ref.~\cite{EV}.

One may think that because of eqs.~(\ref{nllvan}-\ref{nnllvan})
eq.~(\ref{llft}) with $b_0=2 e^{-\gamma_E}$ automatically provides a
result which is correct to N$^2$LL accuracy. This, however, is not
true if $F(L)$ is a physical observable, such as a cross-section.
Indeed, in this case the N$^n$LL approximation to it is
defined by expansion of its logarithm: for example if $F(L)$ is
identified with $\Sigma(\a,\ab\,L)$ eq.~(\ref{ressig}), the expansion
of it to subsequent logarithmic order is given by the expansion
eq.~(\ref{llexp}) of $S(\a,\ab\,L)=\ln\Sigma(\a,\ab\,L) $, and not of
$\Sigma(\a,\ab\,L) $ itself.  The NLL
approximation to the Fourier inverse of $F(L)$ may however be
calculated exactly in terms of $G(L)\equiv \ln F(L)$. One finds
\bea\label{nllfta}
\left[\frac{1}{2\pi}\int d^2\hat b\,e^{-i\hqt\cdot\hat b}\,F(L)\right]_{\rm NLL}
&=&2\,\frac{d}{d\hqtq}\int_0^\infty dx\, J_1(x)\,
\exp\left[G_0+G_1\ln\frac{b_0^2}{x^2}\right]
\nonumber\\\label{nllft}
&=&
2\,\frac{d}{d\hqtq}\,F(\ln\hqtq)\,M\(G'(\ln\hqtq)\),
\eea
where 
\beq
\label{gexpa}
G(L)\equiv\ln F(L)=G_0+G_1\,\ln\frac{b_0^2}{x^2}+O\(\ln^2\frac{b_0^2}{x^2}\),
\eeq
with
\beq
\label{gexpb}
G_0=G(\ln\hqtq);\qquad G_1=G'(\ln\hqtq).
\eeq
This is the result found in ref.~\cite{FNR}. 

\eject

\end{document}